\documentclass[11pt,a4paper]{article}

\usepackage{epsfig,amsmath,amssymb,cite}

\begin{document}
	
\title{Shadow of a charged black hole surrounded by \\ an anisotropic matter field}

\author{Javier Bad\'ia$^{1, 2}$\thanks{e-mail: jbadia@iafe.uba.ar} and Ernesto F. Eiroa$^{1}$\thanks{e-mail: eiroa@iafe.uba.ar}\\
	{\small $^1$ Instituto de Astronom\'{\i}a y F\'{\i}sica del Espacio (IAFE, CONICET-UBA),}\\
	{\small Casilla de Correo 67, Sucursal 28, 1428, Buenos Aires, Argentina}\\
	{\small $^2$ Departamento de F\'{\i}sica, Facultad de Ciencias Exactas y Naturales,} \\ 
	{\small Universidad de Buenos Aires, Ciudad Universitaria Pab. I, 1428, Buenos Aires, Argentina}}
\date{}

\maketitle

\begin{abstract}
	A certain type of matter with anisotropic pressures can add to the Reissner-Nordström metric a term proportional to a power of the radial coordinate. Using the standard method of separating variables for the Hamilton-Jacobi equation, we study the shadow of the corresponding rotating solution, obtained through the Newman-Janis algorithm. We define and calculate three observables in order to characterize the position, size and shape of the shadow.
\end{abstract}

\textit{Keywords}: Black hole shadow; Anisotropic fluid; Electromagnetic field.

\section{Introduction}

There has been a renewed interest in the study of black hole shadows since the release of the first reconstructed image of the surroundings of the supermassive black hole M87* by the Event Horizon Telescope (EHT) collaboration \cite{eht19I,eht19V}. While the shadow of a non-rotating black hole is a circle, a rotating black hole has an asymmetric shadow\cite{bardeen} whose shape depends on the black hole mass and spin, as well as the inclination angle of the observer. In alternative theories of gravity the shadow also presents characteristics that depend on the parameters of the specific model.\cite{perlick21}

Anisotropic fluids in general relativity have been studied in the context of compact objects such as stars or black holes, and in particular spherically symmetric black holes surrounded by an anistropic fluid have recently been introduced\cite{cho18}. The anisotropy allows the matter around the black hole to remain static by means of a negative radial pressure, and the resulting metric has a very general form, also being found in some alternative theories of gravity\cite{kumar20}. In this work we study the shadows of the charged and rotating version\cite{kim20} of these black holes. We adopt units such that $G=c=1$.

\section{Black hole solution}

The energy-momentum tensor corresponding to the anisotropic fluid adopted by Cho \& Kim\cite{cho18} has the form
\begin{equation}
	T_\mu{}^\nu = \mathrm{diag}(-\rho, p_1, p_2, p_2)
\end{equation}
in spherical coordinates, with a radial pressure $p_1$ and an angular pressure $p_2$. A barotropic equation of state
\begin{equation}
	p_i = w_i \rho
\end{equation}
is assumed. It can be shown that in order to have a static black hole solution the value $w_1 = -1$ must be chosen, leaving $w_2$ as the only free parameter, which we will simply name $w$. It is also possible to add to the black hole an electric charge\cite{kiselev03}, leading to the solution
\begin{equation}\label{eq:metrica-esf}
	ds^2 = -f(r)dt^2 + f(r)^{-1}dr^2 + r^2(d\theta^2 + \sin^2\theta\, d\varphi^2),
\end{equation}
with
\begin{gather}
	f(r) = 1 - \frac{2m(r)}{r}, \\
	m(r) = M - \frac{Q^2}{2r} + \frac{K}{2 r^{2w-1}},
\end{gather}
where $M$, $Q$, and $K$ are integration constants. $M$ and $Q$ are the mass and the electric charge of the black hole, respectively, while $K$ is related to the energy density of the anisotropic fluid. It can be seen that by changing the equation of state of the fluid\textemdash that is, by changing $w$\textemdash one arrives at a different power of $r$ in the last term of $f(r)$. The squared charge $Q^2$ is clearly positive; however, there is no impediment to continuing the solution to negative values, replacing $Q^2$ by a parameter $q$ which may take either sign, and we will do so in the following. When $q$ is negative the second term in $m(r)$ can no longer be interpreted as arising from an electric charge, but it can be found for example in some braneworld models\cite{aliev05}.

Applying the Newman-Janis algorithm to the spherically symmetric solution \eqref{eq:metrica-esf} leads to the corresponding axisymmetric metric
\begin{equation}\label{eq:metrica}
	ds^2 = - \frac{\rho^2 \Delta}{\Sigma} dt^2 + \frac{\Sigma \sin^2\theta}{\rho^2} (d\varphi - \Omega\, dt)^2 + \frac{\rho^2}{\Delta} dr^2 + \rho^2 d\theta^2,
\end{equation}
where
\begin{gather}
	\rho^2 = r^2 + a^2 \cos^2\theta, \\
	\Delta = r^2 - 2 m(r) r + a^2, \\
	\Sigma = (r^2 + a^2)^2 - a^2 \Delta \sin^2\theta, \\
	\Omega = \frac{2 a m(r) r}{\Sigma},
\end{gather}
and
\begin{equation}
	m(r) = M - \frac{q}{2r} + \frac{K}{2 r^{2w-1}},
\end{equation}
as in the spherically symmetric case. When $K=0$ the Kerr-Newman metric is recovered, with $q$ replaced by $Q^2$. It is important to keep in mind that outside of general relativity, the metric obtained through the Newman-Janis algorithm may correspond to a different energy-momentum tensor than that of the original spherically symmetric metric\cite{hansen13}. We will assume throughout this work that $w > 0$ so that the spacetime is asymptotically flat. The weak, strong and dominant energy conditions impose various restrictions\cite{cho18,kim20} on the allowed values of $w$ and $K$, but since they do not affect the calculation of the shadow we will not take them into account.

\begin{figure}
	\centering
	\includegraphics[width=\linewidth]{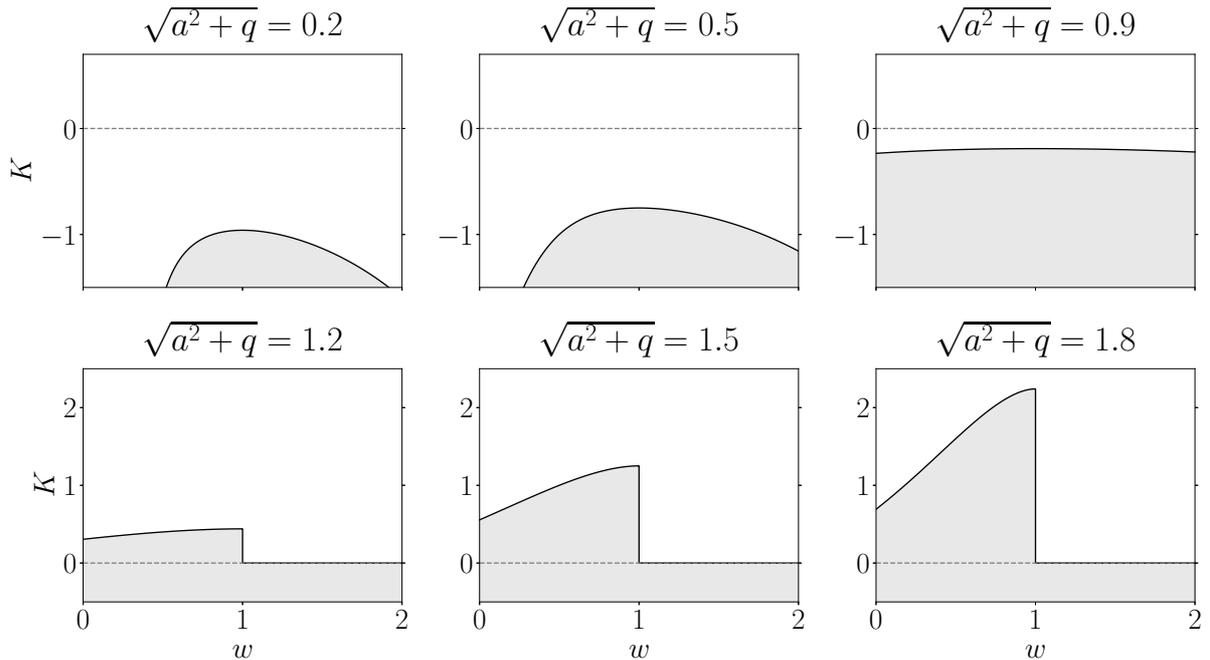}
	\caption{Presence of an event horizon for our adopted metric. The shaded regions indicate the parameter values for which the spacetime contains a naked singularity, while in the white regions there is an event horizon. The solid curves separating them are described paremetrically by Eqs. \eqref{eq:w-crit} and \eqref{eq:k-crit}. We have set $M=1$, so that all parameters are dimensionless.}
	\label{fig:eh}
\end{figure}

We require the presence of an event horizon, so that the spacetime does not contain a naked singularity. The event horizon is located at the largest solution of $\Delta(r) = 0$, so that its disappearance corresponds to parameter values for which $\Delta(r)$ has a double root. The values of $w$ and $K$ for which this happens can be found parametrically as functions of the radius $r$ of the double root\cite{badia20}, and are given by
\begin{equation}\label{eq:w-crit}
	w = \frac{a^2 + q - Mr}{\Delta_\text{KN}}
\end{equation}
and
\begin{equation}\label{eq:k-crit}
	K = \frac{\Delta_\text{KN}}{r^{2r(r-M)/\Delta_\text{KN}}},
\end{equation}
where $\Delta_\text{KN} = r^2 - 2Mr + a^2 + q$ is the functional form of $\Delta$ in a Kerr-Newman-like spacetime. Plotting these curves for various values of $a^2 + q$ shows the regions in parameter space for which an event horizon exists, as seen in Figure \ref{fig:eh}, in which we have set $M=1$ for simplicity. It can be seen from the figure that unlike for the Kerr-Newman spacetime, where $a^2+q \leq M^2$ is a necessary condition for the existence of an event horizon, the presence of the fluid allows for black holes with $a^2+q > M^2$.

\section{The black hole shadow}

We will briefly review the standard method for finding the null geodesics for photons and the shadow, adapted to our selected spacetime\cite{bardeen,badia20}. The metric \eqref{eq:metrica} is independent of the $t$ and $\varphi$ coordinates, leading to the conserved quantities $E = -p_t$ and $L=p_\varphi$; however, there is an additional hidden symmetry with an associated conserved quantity, the Carter constant $\mathcal{Q}$. This constant can be found by assuming a separable solution for the Hamilton-Jacobi equation, leading to the first-order equations of motion
\begin{gather}
	\rho^2 \dot{t} = \frac{r^2+a^2}{\Delta} P(r) + aL - a^2 \sin^2\theta\, E, \\
	\rho^2 \dot{\varphi} = \frac{aP(r)}{\Delta} + \frac{L}{\sin^2\theta} -aE, \\
	(\rho^2 \dot{r})^2 = \mathcal{R}(r), \\
	(\rho^2 \dot{\theta})^2 = \Theta(\theta),
\end{gather}
where
\begin{gather}
	P(r) = E(r^2+a^2) - aL, \\
	\mathcal{R}(r) = P(r)^2 - \Delta [(L-aE)^2 + \mathcal{Q}], \\
	\Theta(\theta) = \mathcal{Q} + \cos^2\theta \left(a^2 E^2 - \frac{L^2}{\sin^2\theta}\right).
\end{gather}
For convenience, we define the impact parameters $\xi = L/E$ and $\eta = \mathcal{Q}/E^2$. The trajectories that make up the shadow contour have the same impact parameters as the spherical photon orbits\textemdash that is, the solutions of the equations of motion that stay at a constant value of $r$. These can be found by solving $\mathcal{R}(r) = 0 = \mathcal{R}'(r)$, and the corresponding impact parameters are given parametrically by
\begin{gather}
	\xi = \frac{4 m(r) r^2 - (r + m(r) + m'(r)r)(r^2 + a^2)}{a(r - m(r) - m'(r)r)}\\
	\eta = r^3 \frac{4a^2 (m(r)-m'(r)r) - r(r - 3m(r) + m'(r)r)^2}{a^2 (r - m(r) - m'(r)r)^2}.
\end{gather}
A distant observer at an inclination angle $\theta = \theta_\text{o}$ can use the celestial coordinates $(\alpha, \beta)$, which are given by
\begin{gather}
	\alpha = - \frac{\xi}{\sin\theta_\text{o}}, \\
	\beta = \pm \sqrt{\eta + \cos^2\theta_\text{o} \left(a^2 - \frac{\xi^2}{\sin^2\theta_\text{o}}\right)}; \label{eq:beta}
\end{gather}
they correspond to horizontal and vertical displacement in the image plane, respectively.

\begin{figure}
	\centering
	\includegraphics[width=\linewidth]{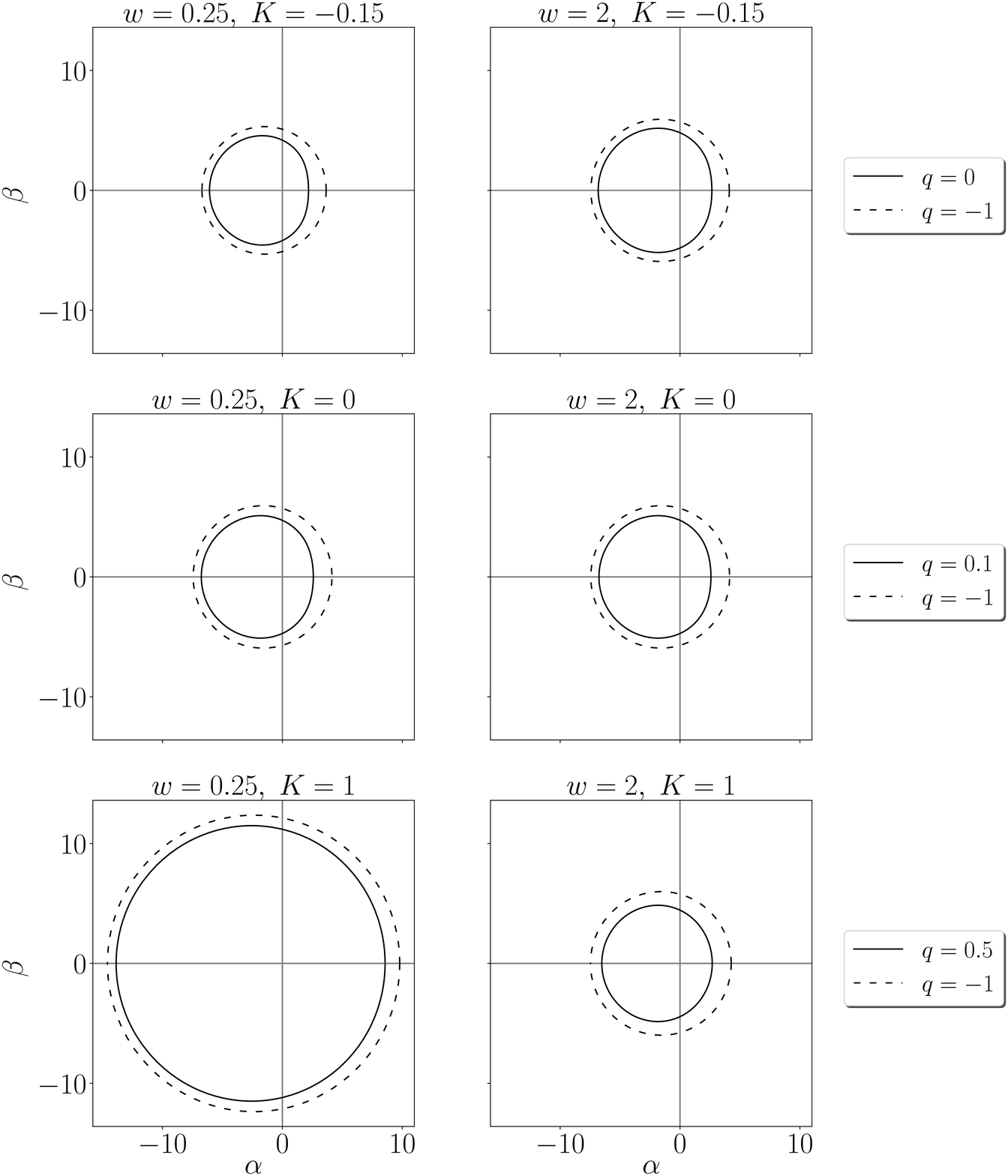}
	\caption{The shadows of a black hole with $a/M = 0.9$ for an observer at $\theta_\text{o} = \pi/2$, for various values of the parameters. All quantities have been adimensionalized by setting $M=1$.}
	\label{fig:sombras}
\end{figure}

Fig. \ref{fig:sombras} shows the shadows produced by a black hole with spin $a/M = 0.9$ as seen by a distant equatorial observer, for various values of the parameters $w$, $K$ and $q$. We find the expected behaviors from the Kerr and Kerr-Newman spacetimes: the shadow is asymmetrical and displaced as a consequence of the spin of the black hole, and its size decreases as the charge increases.

To better characterize the size and shape of the shadow, we define three observables\cite{badia20,kumar20b} that can be calculated from a given shadow contour: the area of the shadow, its oblateness, and its horizontal displacement. The area is simply
\begin{equation}
	A = 2 \int \beta\, d\alpha,
\end{equation}
where the plus sign in Eq. \eqref{eq:beta} has been chosen; the oblateness measures the deviation of the shadow from circularity and is defined as
\begin{equation}
	D = \frac{\Delta\alpha}{\Delta\beta},
\end{equation}
where $\Delta\alpha$ and $\Delta\beta$ are the extent of the shadow in the horizontal and vertical directions respectively; finally, the horizontal displacement is more properly described as the $\alpha$-coordinate of the centroid, given by
\begin{equation}
	\alpha_c = \frac{1}{A} \int 2\alpha\beta\, d\alpha.
\end{equation}

\begin{figure}
	\centering
	\includegraphics[width=\linewidth]{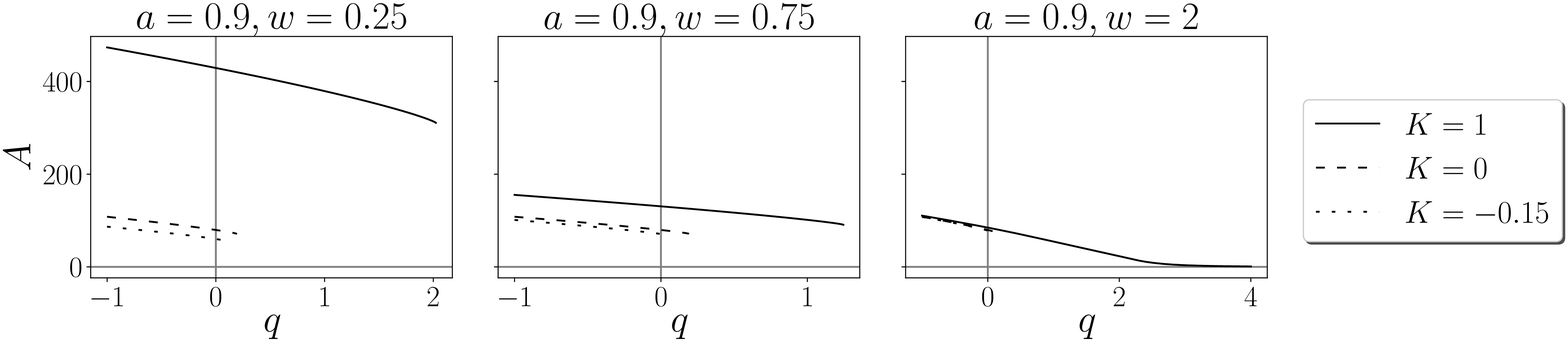}
	\caption{The area of the black hole shadow for some values of the parameters. We set $M=1$, so that all quantities are dimensionless.}
	\label{fig:area}
\end{figure}

\begin{figure}
	\centering
	\includegraphics[width=\linewidth]{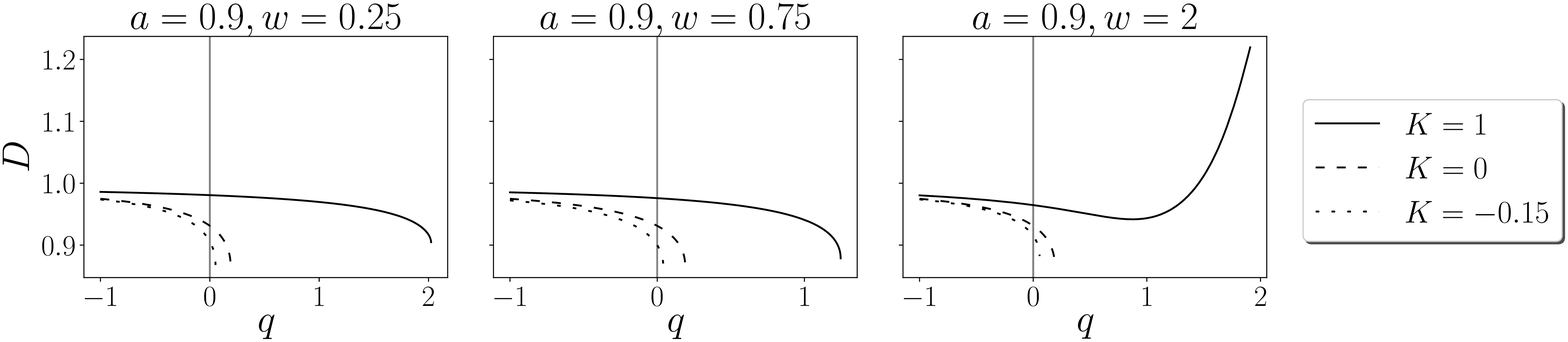}
	\caption{The oblateness of the black hole shadow for some values of the parameters. We set $M=1$.}
	\label{fig:elipt}
\end{figure}

\begin{figure}
	\centering
	\includegraphics[width=\linewidth]{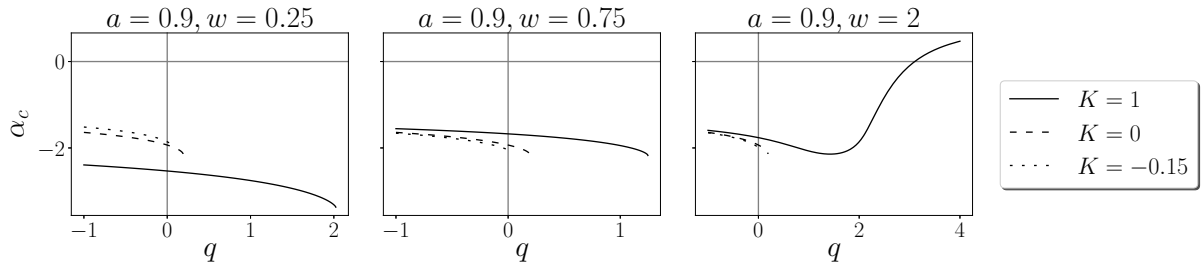}
	\caption{The horizontal position of the shadow centroid for some values of the parameters. We set $M=1$.}
	\label{fig:cent}
\end{figure}

Working with observables has the benefit of making it easier to explore larger areas of parameter space at once; this can be seen in Figs. \ref{fig:area}, \ref{fig:elipt} and \ref{fig:cent}, showing the values of $A$, $D$ and $\alpha_c$ for $a/M = 0.9$ and various values of the other parameters. All plots are shown as functions of $q$, whose maximum value is determined by the requirement that there exist an event horizon.

It is clear from Fig. \ref{fig:area} that the shadow becomes smaller as $q$ increases or as $K$ decreases. This is to be expected, since both parameters appear with opposite signs in the metric. The oblateness and horizontal displacement show more interesting features. For $w < 1$ we see that the shadow becomes more compressed and more displaced as the charge increases. For $w = 2$ and $K > 0$, though, the charge can become large enough to reach a region where the behavior reverses, with $D$ becoming greater than one and $\alpha_c$ becoming greater than zero, reflecting a shadow that is wider than it is tall\textemdash unlike the Kerr shadow\textemdash and displaced in the opposite direction. However, the energy-momentum tensor of the fluid can have the physically undesirable property of having negative energy\cite{kim20,badia20} in this region of parameter space, so the corresponding shadows should be interpreted with care.

\section{Discussion}

In this work we have considered the effect of a fluid with anisotropic pressure on the shadow of a charged and rotating black hole. The fluid makes a contribution to the gravitational field of the black hole, and the resulting Hamilton-Jacobi equation for the geodesics is separable, so that a standard method for finding the shadow contour can be used. We have produced plots of the shadow for various values of the charge as well as the fluid anisotropy and density, and we have found that for many values of the parameters the shadow is qualitatively similar to the well known Kerr-Newman case, exhibiting the typical deviation from circularity due to the spin and shrinking as the charge increases or the energy density of the fluid decreases. However, in some regions of parameter space where the fluid has negative energy, the opposite behavior is seen, with the shadow becoming narrow and displaced in the opposite direction.

\section*{Acknowledgments}

This work has been supported by CONICET and Universidad de Buenos Aires.

\end{document}